\documentclass[a4paper, 12pt]{article}
\usepackage[sort&compress]{natbib}
\bibpunct{(}{)}{;}{a}{}{,} 

\usepackage{algorithm, algorithmicx, algpseudocode}
\usepackage{amsthm, amsmath, amssymb, mathrsfs, multirow, url, subfigure}
\usepackage{graphicx}
\usepackage{ifthen} 
\usepackage{amsfonts}
\usepackage{bbm}
\usepackage[usenames]{color}
\usepackage{fullpage}
\usepackage{csquotes}

\usepackage{bm}

\theoremstyle{plain}

\newtheorem{assumption}{Assumption}
\theoremstyle{definition}

\theoremstyle{remark}

\def\R{\mathbbmss{R}}

\usepackage{authblk}
\author[1]{Indrabati Bhattacharya}
\author[2]{Brent A. Johnson}
\author[3]{William Artman}
\author[4]{Andrew Wilson}
\author[5]{Kevin G. Lynch}
\author[6]{James R. McKay}
\author[7]{Ashkan Ertefaie}
\affil[1,2,3,7]{Department of Biostatistics and Computational Biology, University of Rochester, USA}
\affil[4]{Courant Institute of Mathematical Sciences, New York University, USA}
\affil[5]{Center for Clinical Epidemiology and Biostatistics and Departman of Psychiatry, University of Pennsylvania, USA}
\affil[6]{Department of Psychiatry, University of Pennsylvania, USA}
{
    \makeatletter
    \renewcommand\AB@affilsepx{: \protect\Affilfont}
    \makeatother

    \affil[ ]{Email id}

    \makeatletter
    \renewcommand\AB@affilsepx{, \protect\Affilfont}
    \makeatother

    \affil[1]{indrabati\_bhattacharya@urmc.rochester.edu}
}

\title{A non-parametric Bayesian approach for adjusting partial compliance in sequential decision making}

\date{}

\begin{document}

\maketitle 


\begin{abstract}
Existing methods in estimating the mean outcome under a given dynamic treatment regime rely on intention-to-treat analyses which estimate the effect of following a certain dynamic treatment regime regardless of compliance behavior of patients. There are two major concerns with intention-to-treat analyses: (1)  the estimated effects are often biased toward the null effect; (2) the results are  not generalizable and reproducible due to the potential differential compliance behavior.  These are particularly problematic in settings with high level of non-compliance such as substance use disorder treatments. Our work is motivated by the Adaptive Treatment for Alcohol and Cocaine Dependence study (ENGAGE), which is a multi-stage trial that aimed to construct optimal treatment strategies to engage patients in therapy. Due to the relatively low level of compliance in this trial,  intention-to-treat analyses essentially estimate the effect of being randomized to a certain treatment  sequence which is not of interest. We fill this important gap by defining the target parameter as the mean outcome under a  dynamic treatment regime given potential compliance strata. We propose a flexible non-parametric Bayesian approach, which consists of a Gaussian copula model for the potential compliances, and a Dirichlet process mixture model for the potential outcomes.  Our simulations highlight the need for and usefulness of this approach in practice and illustrate the robustness of our estimator in non-linear and non-Gaussian settings.  

\smallskip

\end{abstract}

\section{Introduction}
\label{s:intro}
A dynamic treatment regime (DTR) is a sequential design that seeks to account for patient heterogeneity based on their response to treatments and their covariate history evolving over time \citep{chakraborty2013statistical, kosorok2015adaptive}. A DTR is a set of decision rules that maps individual characteristics to a treatment option at each decision point. In this paper, we consider estimating the mean outcome under a prespecified DTR in the presence of partial compliance. 

The estimation of the mean outcome under a DTR has been studied extensively,
leading to a wide range of estimators. These estimators include: inverse probability treatment weighted based methods \citep{robins1997causal,  robins2000marginal, murphy2001marginal,van2007causal, bembom2008analyzing,  toh2008causal}; double robust estimators \citep{robins1994estimation, orellana2010dynamic,ertefaie2016identifying}; G-computational formula for dynamic regime \citep{robins1986new}; and targeted maximum likelihood  estimators \citep{rosenblum2010targeted,luedtke2016statistical}. The proceeding methods perform intention-to-treat analyses that ignore the individuals compliance pattern to the assigned treatments. However, in the presence of partial compliance, such analyses only estimate the effect of randomization to a treatment sequence and not the actual treatment effect that is  of main interest. In fact, intention-to-treat analyses are often biased toward the null effect and more importantly not reproducible due to the potential differential compliance behavior in real-world settings \citep{lin2008longitudinal}. To go beyond the intention-to-treat analyses, instrumental variable based methods have  been proposed   for constructing an optimal regime and estimating the corresponding mean outcomes \citep{cui2020semiparametric, qiu2020optimal,qi2021proximal} . However, these approaches apply  only to all-or-none compliance values (i.e., binary) and are not applicable to settings with two (or more) active treatment options at each decision point, which is often the case in pharmacoepidemiologic studies and multi-stage randomized trials \citep{cheng2006bounds, swanson2015selecting, ertefaie2016sensitivity, ertefaie2016selection}. 

Adjusting for partial compliance using standard techniques is subject to  post-treatment stratification bias and lack of causal interpretibility of the results. 
 Principal stratification has been widely used to avoid such problems by classifying individuals into latent classes according to the joint potential compliance values
  instead of adjusting only for the observed ones \citep{frangakis2002principal}. \cite{sjolander2009sensitivity} proposed  to estimate a principal casual effect by dichotomizing the  partial compliance values. This approach, however, can lead to loss of important information due to the dichotomization, and can suffer from the dependence on the threshold of dichotomization. 
\cite{bartolucci2011modeling} proposed a semi-parametric approach; a parametric model for the outcome and a non-parametric copula based model for the intermediate variables. Although, the latter approach does not require dichotomizing the partial compliance values,  the assumption of a parametric family  may not be flexible enough for possibly complex distributions that involves high variability, outliers or skewness \citep{robins2008estimation}. Moreover, the copula method  cannot be trivially extended to accommodate inclusion of covariates. To overcome these issues, \cite{schwartz2011bayesian} proposed a Bayesian semiparametric approach that involves a flexible non-parametric estimation of the joint distribution of the potential compliances using a Dirichlet process mixture (DPM). \cite{kim2019bayesian}  built on their work, and proposed a similar non-parametric Bayesian approach to accommodate multiple post treatment  variables (i.e., mediators) that are measured contemporaneously.

The existing principal stratification based methods  are only applicable in single time point treatment settings (e.g., standard randomized clinical trials). The key challenge in time-varying treatment settings (e.g., multi-stage trials) is that the potential compliance values will also be time-varying. Recently,  \cite{artman2020adjusting} generalized  the semiparametric approach in \cite{schwartz2011bayesian} to a multi-stage randomized trials setting. However, their method is somewhat restrictive. First, the latter assumes a finite dimensional model for conditional expectation of the potential outcome given the potential compliances and the baseline covariates, which may not be flexible enough to capture the true outcome model. Second, their proposed method relies on Gausian error distribution. Third, they imposed some modeling constraints to make the outcome regression coefficients identifiable. 

We propose a non-parametric Bayesian method that extends the existing literature in principal stratification to time-varying treatments and potential compliance settings. A Bayesian approach allows us to treat the missing potential compliances as unknown parameters, and impute them using samples from their posterior distributions. 
A non-parametric approach lends itself to weak assumptions on the identifiability and form of the conditional outcome models, thus 
avoiding the most challenging and cumbersome issues in \citet{artman2020adjusting}. Briefly, our approach for a two-stage setup is summarized as follows. First, we model the marginal distributions of the Stage-1 observed compliances using a DPM, and then use a Gaussian copula to link the marginals to a joint distribution of the Stage-1 potential compliances. Secondly, we use a DPM to model the Stage-2 potential compliances conditionally on the Stage-1 potential compliances, baseline covariates, Stage-1 treatment, and covariates measured just before Stage-2. Then we  specify a flexible DPM model for the joint distribution of the potential outcome, potential compliances and the baseline covariates. Because the complete set of potential compliances are never observed together, we use a data-augmentation approach to impute the unobserved potential compliances based on the joint distribution of the potential compliances. The proposed flexible modeling strategy of the conditional distribution of the potential outcome leads to a robust estimator that does not impose any functional form or distributional assumption on the potential outcome model. 

The rest of the paper is organized as follows. Section \ref{s:engage} describes the motivating dataset this work is based on. Section \ref{s:notation} introduces the basic setup and notations to be used in the paper, and Section \ref{S:modeling} describes the modeling framework. In Section \ref{s: posterior}, we discuss the major steps of the Markov Chain Monte Carlo (MCMC) algorithm used to perform the posterior inference. In Section \ref{s:simulation}, we perform an extensive simulation study to demonstrate the efficacy of our method, and discuss the results. In Section \ref{s:real}, we apply our method on the real data and determine the set of best design embedded dynamic treatment regimes for various compliance values. We close the paper with a brief discussion about the findings of our paper in Section 8.


\section{ENGAGE  study}
\label{s:engage}
This work is motivated by the Adaptive Treatment for Alcohol and Cocaine Dependence (ENGAGE) study \citep{mckay2015effect}. The purpose of this study is to assess the effect of allowing patients who failed to engage in or dropped out of intensive outpatient programs (IOP) to choose their subsequent treatment options. In ENGAGE trial, all patients were first enrolled in the IOP, which required them to attend 3 sessions per week. After two weeks, patients who failed to attend their session at Week 2 were considered non-engaged and they were randomized to one of two motivational interviewing (MI) interventions. Patients in the first group were recommended to engage in IOP (MI-IOP), and the second group was offered a choice for treatment (referred as patient choice or MI-PC). After 8 weeks, the non-engagers (defining as not engaging in any IOP sessions in Weeks 7 and 8) in the first group were re-randomized to either MI+PC or no further care (NFC), while the engagers in the second group received NFC.  

In this study, compliance is defined as the fraction of sessions attended. Thus, there are four potential compliances corresponding to MI-IOP Stage-1, MI-PC Stage-1, MI-PC Stage-2 for MI-IOP Stage-1, and MI-PC Stage-2 for MI-IOP Stage-1, respectively. There are four design embedded  DTRs (EDTRs), and they are defined below:
\begin{enumerate}
    \item Start with MI-IOP. For engagers, offer NFC at the 8-week point, and for non-engagers, offer MI+PC.
    \item Start with MI-IOP. For both engagers and non-engagers at the 8-week point offer NFC.
\end{enumerate}
The other two are similar, with the MI-IOP in the first stage replaced by MI-PC. The aim is to determine the optimal EDTR, which would help the physicians know whether permitting the patients to choose their treatment option is better than leaving it up to the physicians.

\section{Notations, Formulation and Assumptions}
\label{s:notation}

We focus on a two-stage study where binary treatment decisions are made at each time point. 
Suppose we observe $n$ independent, identically distributed trajectories of $$ (\bm{X}_0,A_1,D_{A_1}^{\text{obs}},S,\bm{X}_1,A_2,D_{A_2}^{\text{obs}},Y) \sim P_0.$$
The vector $\bm{X}_0 \in \mathcal{X}_0 \subset \R^{m_1}$ consists of all available baseline covariates measured 
before treatment at the first decision point $A_1\in \{-1,1\}$. Let $S\in \{0,1\}$ be the response indicator to the first stage treatment where $S=1$ for responders.
 The vector $\bm{X}_1 \in \mathcal{X}_1 \subset \R^{m_2}$ 
consists of all available intermediate covariates measured before treatment at the second decision point $A_2\in \{-1,1\}$. We assume that responder individuals (i.e., $S=1$) do not get the second stage treatment (i.e., continue with the first stage treatment). Also, $D_{A_1}^{\text{obs}}\in [0,1]$ and $D_{A_2}^{\text{obs}}\in [0,1]$ represent the level of observed compliance to the  first and the second stage treatments. The observed outcome $Y \in \R$ (measured after $A_2$) is assumed continuous.

Let $D_1$ and $D_2$ denote the potential compliance values under $A_1=1$ and  $A_1=-1$, respectively. Similarly, $D_3$ and $D_4$ denote the potential compliance values under $A_2=1$ for $A_1=1$, and  $A_2=1$ for $A_1=-1$, respectively. Hence, $D_{A_1}^{\text{obs}} = 0.5(A_1+1) D_1 + 0.5(1-A_1)D_2$, and $D_{A_2}^{\text{obs}} = 0.5(A_1+1) D_3 + 0.5(1-A_1)D_4$. In the context of ENGAGE trial, $D_{1}$ and $D_{2}$ denote the potential compliances to MI-IOP Stage-1 $(A_{1}=+1)$ and MI-PC Stage-1 $(A_{1}=-1)$, respectively. The potential compliances to MI-PC Stage-2 $(A_{2}=+1)$ for individuals assigned to MI-IOP Stage-1 $(A_1=+1)$, and MI-PC Stage-1 $(A_1=-1)$ are denoted as $D_{3}$ and $D_{4}$, respectively. We do not consider any potential compliance value for those assigned to NFC at Stage 2 because individuals who are assigned to this treatment option do not have access to the other Stage-2 treatment option.


We also define two types of potential outcomes. Specifically, let $Y_k$ denote the potential outcome of the $k$th treatment sequence, for $k=1,\dots,K$, and $Y^{(l)}$ denote the potential outcome corresponding to the $l$th embedded DTR, for $l=1,\dots,L$. In ENGAGE trial, there are four embedded DTRs (i.e., $L=4$) and  six treatment sequences (i.e., $K=6$), which are described in Table \ref{tab1}.

\begin{table}[h]
\centering
\small
\caption{Observed and latent potential compliances for 6 treatment sequences in ENGAGE study}
\begin{tabular}{cccc} 
$k$ & Treat seq & observed compliance & latent compliance\\
\hline
1 & $A_1=+1, S=1$ & $D_1$ & $D_2$\\
\hline
2 & $A_1=+1, S=0, A_2=+1$ & $D_1,D_3$ & $D_2$\\
\hline
3 & $A_1=+1,S=0, A_2=-1$ & $D_1$ & $D_2,D_3$\\
\hline
4 & $A_1=-1,S=1$ & $D_2$ & $D_1$\\
\hline
5 & $A_1=-1, S=0, A_2=+1$ & $D_2, D_4$ & $D_1$\\
\hline
6 & $A_1=-1, S=0, A_2=-1$ & $D_2$ & $D_1, D_4$
\end{tabular}
\label{tab1}
\end{table}

Our target parameter is the \textit{principal causal effect}, defined as $\text{PCE}^{(l)}(\bm{D},\bm{X}_0)=E(Y^{(l)}\vert \bm{D},\bm{X}_0)$ where $\bm{D}=(D_1,D_2,D_j)$, $j=3,4$.
Using Robins' G-computation formula \citep{robins1986new}, for $l=1,\dots,L$,  the principal causal effects can  be represented as
\begin{align}\label{eq1}
\begin{split}
    \text{PCE}^{(l)}(\bm{D},\bm{X}_0)=&E(Y\vert A_1=a_{1l}, S=1, D_1,D_2,\bm{X}_0)P(S=1\vert A_1=a_{1l},D_1,D_2,\bm{X}_0)\\
    &+E(Y\vert A_1=a_{1l},A_2=a_{2l}, S=0, \bm{D},\bm{X}_0)P(S=0\vert A_1=a_{1l},D_1,D_2,\bm{X}_0),
    \end{split}
\end{align}
where $a_{1l}$ and $a_{2l}$ are the Stage-1 and Stage-2 treatment options that are consistent with the $l$th embedded DTR. The conditional models in (\ref{eq1}) cannot be fit because, for each subject, one or more potential compliances are  latent. We impute the missing potential compliances using the Bayesian non-parametric model described below.

To identify the principal causal effects, we need to impose the following assumptions. 

\begin{assumption}(Stable unit treatment values assumption).
The potential outcome for any unit is not affected by the treatment assignment of other units.
\end{assumption}
\begin{assumption}(Ignorable treatment assignment).
Given the baseline covariates $\bm{X}_0$, the potential outcome and the compliances are independent of the treatment assignments, i.e. $Y^k\perp (A_1,A_2)\vert \bm{X}_0$, $\{D_1,D_2\} \perp A_1 \vert \bm{X}_0$, $\{D_3,D_4\} \perp A_2 \vert \{\bm{X}_0, A_1, D_{A_1}^{obs},\bm{X}_1$\}.
\end{assumption}
\begin{assumption}(Correctly specified copula model).
The joint distribution of the potential compliances conditional on the covariates is a Gaussian copula.
\end{assumption}

Assumption 1 is the no interference assumption which implies that, for a given individual, the potential outcomes and compliance values depend only on the treatment values for that specific individual. Assumption 2 allows us to model the potential outcomes and compliance values using the observed data. Assumption 3 is required to construct the joint distribution of potential compliance values using the observed marginal distributions.

\section{Modeling}
\label{S:modeling}
\subsection{Model for the potential compliances}
We begin by specifying models for the Stage-1 marginal compliances. Because the compliances are between $0$ and $1$, we use a DPM of truncated Gaussian regressions. For marginals of $D_1$ and $D_2$, the DPM is specified in the following way:
\begin{align*}
D_{ji}\vert \bm{X}_{0i}=\bm{x}_{0i}\sim & \int \mathrm{N}(\beta_{0j}+\bm{x}_{0i}^T\bm{\beta}_{j},\sigma_j^2)_{[0,1]}\mathrm{d}G_j(\beta_{0j},\sigma_j^2),\ i=1,\dots,n_j,\ j=1,2,\\
G_j\sim & \mathrm{DP}(\zeta_j,\mathcal{G}_j),
\end{align*}
where $n_j$ is the number of observations for which $D_j$ is observed, and $\mathrm{N}(\mu,\sigma^2)_{[a.b]}$ denotes the normal distribution with mean $\mu$ and variance $\sigma^2$ truncated in $[a,b]$. Note that the mean of the kernel has the form of a linear regression on the baseline covariates $\bm{X}_0$. For simplicity, the mixing is done only over the intercept and variance parameters, and we put a Gaussian prior $\mathrm{N}_P(\bm{\mu}_{\bm{\beta}_j},\Sigma_{\bm{\beta}_j})$ on the regression coefficient parameters $\bm{\beta}_{j}$. The base measure $\mathcal{G}_j$ has been taken as the product of a normal and an inverse gamma distribution, which we can write as $\mathrm{N}(\bm{\mu}_j,S_j)\mathrm{IG}(a_j,b_j)$. 

Next, we model the second stage observed compliances $D_3$ and $D_4$ conditionally on $D_1$, $D_2$, baseline covariates $\bm{X}_0$, and covariates $\bm{X}_1$ measured just before the second stage (i.e., time-varying covariates). We specify the following DPM models for $D_3$ and $D_4$, 
\begin{align*}
    D_{3i}\vert D_{1i}=d_{1i}, \bm{X}_{0i}=\bm{x}_{0i},\bm{X}_{1i}=\bm{x}_{1i}\sim \int\mathrm{N}(\beta_{03}+\gamma_3d_{1i}+\bm{x}_{0i}^T\bm{\beta}_3+\bm{x}_{1i}^T\bm{\beta}_3^\prime, \sigma_3^2)_{[0,1]}\mathrm{d}G_3(\beta_{03},\sigma_3^2),\\
    D_{4i}\vert D_{2i}=d_{2i}, \bm{X}_{0i}=\bm{x}_{0i},\bm{X}_{1i}=\bm{x}_{1i}\sim \int\mathrm{N}(\beta_{04}+\gamma_4d_{2i}+\bm{x}_{0i}^T\bm{\beta}_4+\bm{x}_{1i}^T\bm{\beta}_4^\prime,\sigma_4^2)_{[0,1]}\mathrm{d}G_4(\beta_{04},\sigma_4^2),
\end{align*}
with $G_j\sim \mathrm{DP}(\zeta_j,\mathcal{G}_j)$, $j=3,4$. The treatment sequences for which $D_3$ is observed, also have $D_1$ observed, but $D_2$ latent. Therefore, we only include $D_1$ as a covariate for the marginal model of $D_3$. Similarly, we only include $D_2$, but not $D_1$ for the marginal model of $D_4$.
We consider a normal prior for the regression coefficients $\gamma_3,\gamma_4,\bm{\beta}_3,\bm{\beta}_4,\bm{\beta}_3^\prime$, and $\bm{\beta}_4^\prime$. Finally, the joint distribution of $D_1$, $D_2$, $D_3$ and $D_4$ is modeled by linking the marginals together using a Gaussian copula of the form
\begin{align*}
F_{ D_1,D_2,D_3,D_4}(d_1,d_2,d_3,d_4)=\Phi_{4}[\Phi_1^{-1}\{F_{D_1}(d_1)\},\Phi_1^{-1}\{F_{D_2}(d_2)\},\Phi_1^{-1}\{F_{D_3}(d_3)\},\Phi_1^{-1}\{F_{D_4}(d_4)\}],
\end{align*}
where $F_D$ denotes the distribution function of $D$, and $\Phi_4$ is a 4-dimensional normal distribution with correlation matrix $R$. The joint distribution of the potential compliances cannot be fully identified, because $D_1$ and $D_2$ are never observed together, and neither are $D_3$ and $D_4$. Thus, no amount of data can uniquely estimate the correlation between $(D_1,D_2)$ and $(D_3,D_4)$. The information about the correlation between potential compliances is implicitly embedded in the $Y$ model where the potential compliances appear.  This is evident in the compliance augmentation step in Section \ref{s: posterior}, where the distribution of the missing compliances are retrieved from  the distribution of $Y$.  The association information of $(D_1,D_2,D_3,D_4)$ and the $Y$ model can be complex, which highlights the importance of the proposed flexible Bayesian non-parametric model for potential compliances via the DPM models. 

Moreover, inference on $R$ needs careful attention because $R$ has to be positive definite. To provide an inference for the elements of $R$, we put a $\mathrm{U}(-1,1)$ prior on each of the off-diagonal elements in $R$. For posterior inference on $R$, we use a Metropolis-Hastings algorithm, and we choose a proposal distribution that respects this restriction. Details are described in Section \ref{s: posterior}.

\subsection{Model for the potential outcome}
To model the potential outcomes given the potential compliances and the baseline covariates for each of the six treatment sequences, we use a locally weighted mixture of Gaussian regression models by specifying a DPM model for the joint distribution of the potential outcome, potential compliances and the baseline covariates. The potential outcomes corresponding to sequences 2 and 3 may depend on $(D_1,D_2,D_3)$, and the potential outcomes corresponding to sequences 5 and 6 may depend on $(D_1,D_2,D_4)$.
It is natural to assume that the potential outcomes among responders (sequences 1 and 4) to not depend on the second stage compliances $D_3$ and $D_4$. 

For responders, the conditional distribution of the potential outcome $Y_i$ is specified as
\begin{align*}
f_Y(y_i^k\vert D_{1i}=d_{1i},D_{2i}=d_{2i},\bm{X}_{0i}=\bm{x}_{0i})=
\sum_{b=1}^\infty \psi_b^k\mathrm{N}(y_i,h_{1i},h_{2i},\bm{x}_{0i} \vert \mu_b^k,\Sigma_b^k),
\end{align*}
where $h_{ji}=\Phi_1^{-1}\{F_{D_j}(d_{ji})\}$, and $\psi_b^k=\xi_b^k/\sum_{l=1}^\infty \xi_l^k \mathrm{N}(H_{1i}=h_{1i},H_{2i}=h_{2i},\bm{X}_{0i}=\bm{x}_{0i}\vert \mu_{l,-1}^k,\\ \Sigma_{l,-1,-1}^k)$, for $k=1,4$. Here $\mu_{l,-1}^k$ denotes all elements of the mean vector $\mu_l^k$ except for $Y_i$ and $\Sigma_{l,-1,-1}^k$ denotes the submatrix of the joint covariance matrix $\Sigma_l^k$ formed by deleting the first row and first column. Moreover, $\xi_b^k={\xi_b^\prime}^k\prod_{h<b}(1-{\xi_h^\prime}^k)$ are the stick-breaking weights constructed from ${\xi_b^\prime}^k \sim \mathrm{Beta}(1,\alpha)$, for some $\alpha>0$. The definition of $\bm{H}=(H_1,H_2)$ justifies the use of a mixture of multivariate Gaussians for the joint distribution of $(Y,\bm{H},\bm{X}_0)$. For sequences 1 and 4, $D_2$ and $D_1$ have to be imputed, respectively. 

For $k=2,3$, i.e., for non-responders who were assigned to MI-IOP in Stage-1, the conditional distribution of the potential outcomes is modeled as
\begin{align*}
f_Y(y_i^k\vert D_{1i}=d_{1i},D_{2i}=d_{2i},D_{3i}=d_{3i},\bm{X}_{0i}=\bm{x}_{0i})=\sum_{b=1}^\infty \psi_b^k\mathrm{N}(y_i,h_{1i},h_{2i},h_{3i},\bm{x}_{0i}\vert \mu_b^k,\Sigma_b^k),
\end{align*}
with the notations carrying the same meaning. For the remaining sequences, i.e., $k=5,6$, we model the potential outcomes in a similar fashion, 
\begin{align*}
f_Y(y_i^k\vert D_{1i}=d_{1i},D_{2i}=d_{2i},D_{4i}=d_{4i},\bm{X}_{0i}=\bm{x}_{0i})=\sum_{b=1}^\infty \psi_b^k\mathrm{N}(y_i,h_{1i},h_{2i},h_{4i},x_{0i}\vert \mu_b^k,\Sigma_b^k).
\end{align*}
Notice that, we do not include the time-varying covariates $\bm{X}_1$ in neither of the outcome models. The reason behind this is that in the definition of PCE in \eqref{eq1}, we are only interested in the conditional expectation of the causal effect given a principal stratum, not in the causal effect of the stratum itself. Thus, we should not adjust for the time-varying covariates in the outcome models.

We use a Markov Chain Monte Carlo (MCMC) algorithm for the posterior inference. Imputation of the missing potential compliances and estimation of the potential compliances are connected together in each MCMC iteration (described in detail in Section \ref{s: posterior}).  
\subsection{Stage-1 response indicator model}
We model $P(S=1\vert A_1,D_1,D_2,\bm{X}_0)$, i.e., the response probabilities to Stage-1 treatments, using  a Bayesian non-parametric approach; i.e., a Dirichlet process mixture of logistic kernels. For $A_1=1$,we specify the following DPM model,
\begin{align*}
    P(S_i=1\vert A_{1i}=1,D_{1i}=d_{1i},D_{2i}=d_{2i},\bm{X}_{0i}=\bm{x}_{0i})=\sum_{j=1}^\infty\omega^{(1)}_j\delta_{\lambda_{ij}^{(1)}},
\end{align*}
where $\lambda^{(1)}_{ij} \sim \mathrm{Bernoulli}(p^{(1)}_{ij})$, with $\text{logit}(p^{(1)}_{ij})=\alpha^{(1)}_{j}+\beta^{(1)}_{1j}d_{1i}+\beta^{(1)}_{2j}d_{2i}+\bm{x}_{0i}^T\bm{\gamma}_{j}^{(1)}$, $i=1,\dots,N_1$ and $\omega^{(1)}_{j}=\omega_{jl}^\prime\prod_{m=1}^{j-1}(1-\omega_{ml}^\prime)$ are the stick-breaking weights constructed from $\omega_{ml}^\prime\overset{iid}{\sim}\mathrm{Beta}(1,\alpha)$. Here, $N_1$ denotes the number of observations with $A_1=1$. Also, $\alpha_1^{(1)},\alpha_2^{(1)},\dots, \overset{iid}{\sim}\mathrm{N}(1,1)$, $\beta^{(1)}_{11},\beta^{(1)}_{12},\dots\overset{iid}{\sim}\mathrm{N}(1,1)$, $\beta^{(2)}_{11},\beta^{(2)}_{12},\dots\overset{iid}{\sim} \mathrm{N}(1,1)$, and $\bm{\gamma}_1^{(1)},\bm{\gamma}_2^{(1)},\dots\overset{iid}{\sim} \mathrm{N}_{m_1}(\mathbf{1}_{m_1},\mathbf{I}_{m_1})$, where $\mathbf{c}_k$ denotes the vector of length $k$, and $\mathbf{I}_{k\times k}$ denotes the identity matrix of order $k\times k$. Similarly, for $A_1=-1$, we model the response probability as
\begin{align*}
    P(S_i=1\vert A_{1i}=-1,D_{1i}=d_{1i},D_{2i}=d_{2i},\bm{X}_{0i}=\bm{x}_{0i})=\sum_{j=1}^\infty\omega^{(-1)}_j\delta_{\lambda_{ij}^{(-1)}},
\end{align*}
with $\lambda_{ij}^{(-1)}\sim \mathrm{Bernoulli}(p_{ij}^{(-1)})$, and $\text{logit}(p^{(-1)}_{ij})=\alpha^{(-1)}_{j}+\beta^{(-1)}_{1j}d_{1i}+\beta^{(-1)}_{2j}d_{2i}+\bm{x}_{0i}^T\gamma_{j}^{(-1)}$, $i=1,\dots,N_2$, where $N_2$ is the number of observations for which $A_1=-1$. The stick-breaking weights $\omega_j^{(-1)}$ are constructed in a similar fashion as before, and $\alpha_j^{(-1)},\beta_{1j}^{(-1)},\beta^{(-1)}_{2j}$ and $\gamma_j^{(-1)}$ for $j=1,2,\dots$ are drawn from Gaussian distributions as before. 

The posterior inference for the response indicator model is done separately from the potential outcome estimation, and then are plugged in \eqref{eq1} for estimation of the EDTR. Note that, for $A_1=+1$, $D_2$'s are unobserved, and for $A_1=-1$, $D_1$'s are unobserved. We've already imputed the missing potential compliances in course of the estimation of potential compliances, and we use those values for estimation of the response probabilities as well.

\section{Posterior inference}
\label{s: posterior}
The posterior inference consists of five major steps. The posterior inference for the DPM models using the truncation approximation \citep{ishwaran2001gibbs}, i.e., we truncate the stick-breaking weights at a large number $N$. One can choose the truncation level according to Theorem 2 in \cite{ishwaran2001gibbs}, which says that, for a $\mathrm{DP}(MG)$ process, 
$$
\Vert \mu_N-\mu_\infty\Vert_\infty\sim 4n\ \mathrm{exp}(-(N-1)/M),
$$
where $\mu_N$ denotes the marginal density under the truncation $N$, and $\mu_\infty$ denotes its limit, with $n$ being the sample size. We follow \citep{kim2019bayesian} for hyperprior specifications.
\begin{enumerate}
\item[Step (1)] {\it Sampling parameters for the marginals of the Stage-1 observed  compliances (i.e., $D_1$ and $D_2$).} We make use of the stick-breaking representation of the DPM models. The cluster weights and the latent cluster variables are updated using the standard Block Gibbs sampler for DPM models \citep{ishwaran2001gibbs}. The truncated normal kernel makes it difficult to sample the  cluster-specific parameters $(\beta_{0j},\sigma_j^2),\ j=1,2$ using the standard Gibbs sampler, hence we use block Metropolis steps instead. For the marginal distributions of the Stage-1 observed compliances, we sample the necessary parameters and update $H_{1i}=\Phi_1^{-1}\{F_{D_1}(d_{1i};\bm{\theta}_1,\bm{x}_{0i})\}$, $H_{2i}=\Phi_1^{-1}\{F_{D_2}(d_{2i};\bm{\theta}_2,\bm{x}_{0i})\}$, $i=1,\dots,n_j$,  where $\theta_j$ denotes the vector of parameters corresponding to the $j$th marginal, for $j=1,2$. The likelihood of $\bm{\theta}=(\bm{\theta}_1,\bm{\theta}_2,\bm{\theta}_3,\bm{\theta}_4)$ and $R$ in the copula model has the form
\begin{align*}
    f(\bm{\theta}, R)=\prod_{i=1}^n\bigg[\vert & R\vert^{-1/2}\exp\bigg\{-\frac{1}{2}{\bm{d}}^T_{i}(1-R^{-1})\bm{d}_{i}\bigg\}\times f_{D_1}(d_{1i}\vert \bm{\theta}_1,\bm{x}_{0i})f_{D_2}(d_{2i}\vert \bm{\theta}_2,\bm{x}_{0i})\\
    &\times f_{D_3}(d_{3i}\vert \bm{\theta}_3,\bm{x}_{0i},d_{1i},\bm{x}_{1i})f_{D_4}(d_{4i}\vert \bm{\theta}_4,\bm{x}_{0i},d_{2i},\bm{x}_{1i})\bigg],
\end{align*}
where $\bm{d}_i=(d_{1i},d_{2i},d_{3i},d_{4i})$. Basically, Step 1 consists of sampling from $f(\bm{\theta}_j\vert \bm{\theta}_{-j},d_j,\\ \bm{x}_0,R)$ for $j=1,2$. We obtain the conditional likelihood as
\begin{equation}\label{eqs1}
\begin{split}
    \log f(\bm{\theta}_j\vert d_j,\bm{\theta}_{-j},\bm{x}_0,R)=&\text{const}+\frac{1}{2}(1-R_{jj}^{-1})\sum_{i=1}^n h_{ji}^2-\sum_{i=1}^n\sum_{k\neq j}(R^{-1})_{jk}h_{ji}h_{ki}\\
    &+\sum_{i=1}^n\log f_{D_j}(d_{ji}\vert \bm{x}_{0i},\bm{\theta}_j)+\log f(\bm{\theta}_j),
    \end{split}
\end{equation}
with $h_{ji}=\Phi^{-1}\{F_{d_j}(d_{ji},\bm{\theta}_j,\bm{x}_{0i})\}$. In Step 1, we update $\bm{\theta}_1$ and $\bm{\theta}_2$ from $f(\bm{\theta}_1\vert d_1, \bm{\theta}_{2},\bm{x}_0,\\ R)$ and $f(\bm{\theta}_2\vert d_2, \bm{\theta}_{1},\bm{x}_0,R)$ respectively. The marginal models for $D_1$ and $D_2$ have the following truncated stick-breaking representation
\begin{align*}
    f_{D_j}(d_{ji}\vert \bm{\theta}_j)=\sum_{b=1}^B \omega_{j(b)}  \mathrm{N}(d_{ji};\beta_{j0(b)}+\bm{x}_{0i}^T\bm{\beta}_{j},\sigma_{j(b)}^2)_{[0,1]},
\end{align*}
where $\omega_{j(b)}=\omega_{j(b)}^\prime\prod_{h<b}(1-\omega_{j(b)}^\prime)$, are the stick-breaking weights constructed from $\omega_{j(b)}^\prime \sim \mathrm{Be}(1,\lambda_j)$, $j=1,2$, and $B$ is the number of clusters. Also, $(\beta_{0j(b)},\sigma_{j(b)}^2)\overset{iid}{\sim}\mathrm{N}(\mu_j,S_j)\mathrm{IG}(a_j,b_j)$, $b=1,\dots,B$. Following the prior specifications in \cite{kim2019bayesian}, the hyperpriors are specified as follows: $\lambda_j\sim \mathrm{G}(1,1)$, $\mu_j\sim\mathrm{N}(\mu_j^\star,S_j^\star)$, $S_j\sim\mathrm{G}(a_j^\star,b_j^\star)$, with $a^\star_j\sim\mathrm{Unif}(1,5)$, and $b_j^\star=100a_j^\star$. The truncation level $B$ has been chosen to be 8. We choose $\mu_j^\star=\bar{D}_j$, and $S_j^\star=1$, and also $a_j=b_j=1$. The sub-steps are described below:
\begin{itemize}
    \item Denote the latent cluster variable for the subject $i$ as $Z_i$, which takes value in $\{1,2,\dots,B\}$. At the $t$th iteration, denote the values of $\beta_{j0(b)},\sigma_{j(b)}^2$, and $\omega_{j(b)}$ as $\beta_{j0(b)}(t),\sigma_{j(b)}^2(t)$, and $\omega_{j(b)}(t)$ respectively, for $b\in\{1,\dots,B\}$. Draw $Z_i$ from $Z_i\sim \text{Categorical}(p_1,\dots,p_B)$, where $p_b=\mathrm{N}\{D_{ji};\beta_{j0(b)}(t-1),\sigma_{j(b)}^2(t-1)\}\times \omega_{j(b)}(t-1)$. Draw $\omega_{j(b)}^\prime\sim\mathrm{Be}(1+m_b,\lambda_j+\sum_{q=1}^{B-1}m_q)$, where $m_b$ is the number of subjects having $Z_i=b$. Next, we update $\omega_{j(b)}$ via $\omega_{j(b)}=\omega_{j(b)}^\prime\prod_{h<b}(1-\omega_{h(b)}^\prime)$. Finally, we update $\lambda_j\sim \mathrm{Ga}(B,1-\sum_{b=1}^B\log(1-\omega_{j(b)}))$.
    \item Update $a_j^\star,\mu_j,S_j$: Denote the values at $t$th iteration as $a_j^\star(t),\mu_j(t),S_j(t)$. The proposal distributions are as follows: $a^2_{j(\text{prop})}\sim \mathrm{Unif}(1,5)$, $\mu_{j(\text{prop)}}\sim\mathrm{N}(\mu_j(t-1),1)$, $S_{j(prop)}\sim \mathrm{Unif}(S_{j(prop)}(t-1)-0.1,S_{j(prop)}(t-1)+0.1)$ and denote the joint distribution as $q\{\Psi_{prop};\Psi(t-1)\}$. Then the acceptance probability of the proposed values is given by
        \begin{align*}
            AR_1=\mathrm{min}\Big\{1,\frac{f(\bm{\theta}_{j(prop)}\vert d_j,h_{-j},\bm{x}_0,R)q\{\Psi_{prop};\Psi(t-1)\}}{f(\bm{\theta}_{j}(t-1)\vert d_j,h_{-j},\bm{x}_0,R)q\{\Psi(t-1);\Psi_{prop}\}}\Big\},
        \end{align*}
        where $f(\bm{\theta}_{j\text(prop)}\vert d_j,h_{-j},\bm{x}_0,R)$ denotes the conditional distribution in Equation \eqref{eqs1} before the log transformation. 
        \item Update $\beta_{j0(b)}$, $b=1,\dots,B$: propose values from $\beta_{j0(b)(prop)}\sim N\{\beta_{j0(b)}(t-1),0.1\}$ and use the same metropolis step as in the previous step.
        \item Update $\sigma^2_{j(b)},b=1,\dots,B$: propose values from $1/\sigma^2_{j(b)}\sim G\{0.1/\sigma^2_{j(b)}(t-1)\times c,c\}$ for $b=1,\dots,B$ where $c$ is a large constant to concentrate the probability around $1/\sigma^2_{j(b)}(t-1)$. Here we choose $c=15$.
        \item Update $\bm{\beta}_j=(\beta_{j1},\dots,\beta_{jm_1})$: Following  \cite{kim2019bayesian}, we propose values according to $\bm{\beta}_{j(\text{prop})}\sim \mathrm{N}_{m_1}\{\bm{\beta}_j(t-1),\Sigma^\prime_{\bm{\beta}_j}\}$, with $\Sigma^\prime_{\bm{\beta}_j}=\hat{\Sigma}_{\bm{\beta}_j}+\bm{I}_{m_1}$, where $\hat{\Sigma}_{\bm{\beta}_j}$ is the covariance matrix of the regression coefficients of $D_j$ on $\bm{X}_0$, and $\bm{I}_N$ denotes the identity matrix of order $N\times N$. We use the same metropolis sub-step used for the update of $\beta_{j0(b)},$ $b=1,\dots,B$.
\end{itemize}
\item[Step (2)] {\it Updating the parameters for marginals of $D_3$ and $D_4$, the second stage observed compliances.} Similar to Step 1, with the only difference being that we have more covariates for the marginal regression models, namely, the first stage observed compliances and the covariates measured just before the second stage of treatment assignment. We update the parameters for the marginal distributions of $D_3$ and $D_4$, and update $H_{3i}=\Phi_1^{-1}\{F_{D_3}(d_{3i};\bm{\theta}_3,\bm{x}_{0i},d_{1i},\bm{x}_{1i})\}$ and $H_{4i}=\Phi_1^{-1}\{F_{D_4}(d_{4i};\bm{\theta}_4,\bm{x}_{0i},d_{2i},\\\bm{x}_{1i})\}$. the conditional likelihood for $\bm{\theta}_j,\ j=3,4$ can be written as
\begin{equation}\label{eq2}
\begin{split}
    \log f(&\bm{\theta}_3 \vert d_3,d_1,\bm{\theta}_{-3},\bm{x}_0,\bm{x}_1,R)=\text{const}+\frac{1}{2}(1-R_{33}^{-1})\sum_{i=1}^n h_{3i}^2-\\
    &\sum_{i=1}^n\sum_{k\neq 3}(R^{-1})_{3k}h_{3i}h_{ki}+\sum_{i=1}^n\log f_{D_3}(d_{3i}\vert \bm{x}_{0i},d_{1i},\bm{x}_{1i},\bm{\theta}_3)+\log f(\bm{\theta}_3),
    \end{split}
\end{equation}
and 
\begin{equation}\label{eq3}
\begin{split}
    \log f(&\bm{\theta}_4 \vert d_4,d_2,\bm{\theta}_{-4},\bm{x}_0,\bm{x}_1,R)=\text{const}+\frac{1}{2}(1-R_{44}^{-1})\sum_{i=1}^n h_{4i}^2-\\
    &\sum_{i=1}^n\sum_{k\neq 4}(R^{-1})_{4k}h_{4i}h_{ki}+\sum_{i=1}^n\log f_{D_4}(d_{4i}\vert \bm{x}_{0i},d_{2i},\bm{x}_{1i},\bm{\theta}_4)+\log f(\bm{\theta}_4).
    \end{split}
\end{equation}
The marginals of $D_3$ and $D_4$ have a similar stick-breaking representation:
\begin{align*}
    f_{D_3}(d_{3i}\vert \theta_3)=&\sum_{b=1}^B \omega_{3(b)}  \mathrm{N}(d_{3i};\beta_{30(b)}+\bm{x}_{0i}^T\bm{\beta}_{3}+\gamma_3d_{1i}+\bm{x}_{1i}^T\bm{\beta}^\prime_3,\sigma_{3(b)}^2)_{[0,1]},\\
    f_{D_4}(d_{4i}\vert \theta_4)=&\sum_{b=1}^B \omega_{4(b)}  \mathrm{N}(d_{4i};\beta_{40(b)}+\bm{x}_{0i}^T\bm{\beta}_{4}+\gamma_4d_{2i}+\bm{x}_{1i}^T\bm{\beta}^\prime_4,\sigma_{4(b)}^2)_{[0,1]}.
\end{align*}
We update $w_{j(b)},\beta_{j(b)},\sigma_{j(b)}^2,\bm{\beta}_j$, $b=1,\dots,B$, $j=3,4$, in the same way we did before, so we omit the details. The only extra sub-steps are updating the regression coefficients corresponding to the time-varying covariates $\bm{X_1}$ and Stage-1 observed compliances $D_1$ and $D_2$, which we describe below.
\begin{itemize}
\item Update $(\bm{\beta}_j,\bm{\beta}^\prime_j,\gamma_j)$, $j=3,4$: We propose values according to $(\bm{\beta}_{j(\text{prop})},\ \bm{\beta}^\prime_{j(\text{prop})},\\ \gamma_{j(\text{prop})})\sim \mathrm{N}_{m_1+m_2+1}(\{\bm{\beta}_j(t-1),\bm{\beta}^\prime_j(t-1),\gamma_j(t-1)\},\Sigma^\prime_{\bm{\beta}_j,\bm{\beta}^\prime_j,\gamma_j})$, with $\Sigma^\prime_{\bm{\beta}_j,\bm{\beta}^\prime_j,\gamma_j}=\hat{\Sigma}_{\bm{\beta}_j,\bm{\beta}^\prime_j,\gamma_j}+\bm{I}_{m_1+m_2+1}$, where $\hat{\Sigma}_{\bm{\beta}_j,\bm{\beta}^\prime_j,\gamma_j}$ is the empirical covariance matrix of the regression coefficients of $D_j$ on $\bm{X}_{0},D_{1},\bm{X}_{1}$ for $j=3$, and $\bm{X}_{0},D_{2},\bm{X}_{1}$ for $j=4$. We use the same metropolis sub-step used for the update of $\beta_{j0(b)},$ $b=1,\dots,B$ in Step 1.
\end{itemize}
\item[Step (3)] {\it Updating the copula correlation matrix $R$.} As already mentioned, we perform the inference using a uniform prior on the off-diagonal elements. The posterior sampling is done using  Metropolis steps. propose each association parameter $r_{(prop)}\sim \mathrm{Unif}(r_{L},r_{U})$ where $r_{L}$ and $r_{U}$ are determined to give the positive definite matrix $R$. The acceptance probability of the proposed values is then given by
    \begin{align*}
        AR_2=\min\bigg\{1,\frac{f(R_{r(prop)}\vert \bm{d},\bm{x}_0,\bm{x}_1,\theta)q\{r(t-1)\}}{f(R_{r}(t-1)\vert \bm{d},\bm{x}_0,\bm{x}_1,\theta)q(r_{(prop)})}\bigg\},
    \end{align*}
    where $R_{r(prop)}$ denotes the correlation matrix with the $r$th element set to the proposed value, and all the other entries set to their current values. Similarly, $R_{r}(t-1)$ denotes the correlation matrix with the $r$th element set to the value at the $(t-1)$th iteration, and all the other entries set to their current values.
    
    Next, we briefly describe how to choose $r_L$ and $r_U$. Suppose, we start with a positive definite matrix $R$, and we define $R(r)$ to be the matrix obtained by changing the $(i,j)$th element to $r$. As pointed out in \cite{barnard2000modeling}, $R(r)$ is positive definite if and only if $f(r)=\vert R(r)\vert>0$, and $f(r)$ is a quadratic function in $r$. Hence, we can find the interval for $r$ that leads to $R$ being positive definite by solving the quadratic equation $f(r)=0$.
\item[Step (4)] {\it Augmenting the missing compliances.} We use standard Metropolis steps, with the proposal distribution being a truncated normal with its mean set at the current value. Draw $D_{j(prop)}\sim \mathrm{N}(D_j(t-1),s_{D_j})_{[0,1]}$ and calculate acceptance probability as follows:
    \begin{align*}
        AR_3=\min\bigg\{1,\frac{f(d^{ \text{mis}}_{j(\text{prop})},d_j^{\text{obs}},d_{- j},\bm{x}_0,\bm{x}_1\vert \bm{\theta},R)q\{d_j(t-1)\}}{f(d^{\text{mis}}_{j}(t-1),d_j^{\text{obs}},d_{-j},\bm{x}_0,\bm{x}_1\vert \bm{\theta},R)q(d_{j(prop)})}\bigg\},
    \end{align*}
    where $s_D$ is the empirical standard deviation of $D$.
\item[Step (5)] {\it Estimating the potential outcome using the DPM model.} Once we have the missing potential compliances, we use the locally weighted DPM model to get the posterior estimate of the potential outcomes. For $k=1,4$, we obtain posterior samples from the conditional distribution of $Y^k$ given $\bm{X}_0$, $D_1$ and $D_2$. For $k=2,3$, we draw samples from the posterior conditional distribution of $Y^k$ given $\bm{X}_0,D_1$, $D_2$, and $D_3$. Similarly, for $k=5,6$, we obtain samples from the conditional distribution of $Y^k$ given $\bm{X}_0,D_1,D_2$, and $D_4$. We use the truncation approximation and the block Gibbs sampler for DPM models to estimate the conditional density. The steps are straightforward, and hence omitted.
\end{enumerate}
We draw 10000 samples from the posterior, with a thinning of 5, which gives us 2000 many posterior samples. The trace plots of the MCMC chains show evidence of convergence. The R codes used to obtain the results are available at the Github repository \url{https://github.com/indrabati646/Partial-compliance}.


\section{Simulation studies}
\label{s:simulation}
\subsection{Data generative models}
We assess the practical performance of our proposed principal causal effect 
estimator in various settings including linear and non-linear outcome models with Gaussian errors, and linear outcome models with non-Gaussian errors. 

In scenarios 1 and 2, $A_1 \sim \text{Bernoulli}(0.5)$, $A_2 \sim \text{Bernoulli}(0.5)$, baseline covariates $X_{01}\sim N(-0.5,0.3^2)$, $X_{02}\sim N(0,0.1^2)$, $X_{03}\sim\mathrm{N}(0.5,0.3^2)$, the Stage-1 response indicators are generated as $S_i \mid D_1, X_{0},A_1=+1\sim \mathrm{Bern}\Big\{\frac{\mathrm{exp(D_{1}-1.5+0.2X_{02})}}{1+\mathrm{exp(D_{1}-1.5+0.2X_{02})}}\Big\}$, and $S_i \mid D_1, X_{0},A_1=-1 \sim \mathrm{Bern}\Big\{\frac{\mathrm{exp(D_{1}-1.5+0.3X_{03})}}{1+\mathrm{exp(D_{1}-1.5+0.3X_{03})}}\Big\}$. For scenario 3, we have two baseline covariates $X_{01}\sim N(-0.5,0.3^2)$, $X_{02}\sim N(0,0.1^2)$, and a time-varying covariate $X_{11}\sim N(0.5+0.3X_{01}+0.7X_{02}+0.1A_1,0.1^2)$ . Also, in scenarios 1 and 2,  the first stage treatment does not have an effect on the second stage potential compliances, thereby,  $D_3=D_4$. In scenario 3, we relax this restriction.

In each setting, we
sample $n \in \{250,500,1000 \}$ independent and identically
distributed observations, applying the proposed estimator and the one in \cite{artman2020adjusting} to the resultant data. This
was repeated $200$ times.

\noindent \textbf{Scenario 1 (Linear models with Gaussian errors).} We generate the marginal observed compliances, $D_1$, $D_2$ and $D_3$ from a truncated normal with means $0.5X_{01}+0.5X_{02}$, $0.5X_{02}$, and $0.5X_{03}-0.5X_{01}$ respectively, and variance 0.25. The joint distribution of $D_1,D_2,D_3$ is specified by a Gaussian copula with the correlation matrix
\begin{equation*}
R^\star=\begin{pmatrix}
    1 & 0.2 & 0.2\\
    0.2 & 1 & 0.2\\
    0.2 & 0.2 & 1
    \end{pmatrix}.
\end{equation*}

The generative model results in $K=6$ treatment sequences. We generate the outcomes corresponding to each sequence using the following models \citep{artman2020adjusting}:
\begin{align*}
    Y_{1i}=& 0.7+0.6D_{1i}+0.8X_{01i}-0.2X_{02i}+\epsilon_{1i},\\ 
    Y_{2i}=& 0.2+0.7D_{1i}+0.9D_{3i}+0.4D_{1i}D_{3i}-0.9X_{01i}+0.6X_{03i}+\epsilon_{2i},\\
    Y_{3i}=& 0.2+ 0.6D_{1i}+0.9D_{3i}+0.4D_{1i}D_{3i}-0.9X_{01i}+0.6X_{03i}+\epsilon_{3i},\\
    Y_{4i}=& 0.7+0.6D_{1i}+0.6D_{2i}+0.8X_{01i}-0.2X_{02i}+\epsilon_{4i},\\
    Y_{5i}=& 0.3+0.6D_{2i}+0.7D_{3i}+0.7D_{2i}D_{3i}-0.5X_{02i}+\epsilon_{5i},\\
    Y_{6i}=& 0.3+0.8D_{2i}+0.7D_{3i}+0.7D_{2i}D_{3i}-0.5X_{02i}+\epsilon_{6i},
\end{align*}
with $\epsilon_{ki}\sim N(0,0.1^2)$, $k=1,\dots,6$.

\noindent \textbf{Scenario 2 (Linear models with non-Gaussian errors).} In this scenario, we study the performance of our proposed method under a true error distribution that is more complex (i.e., bimodal) than Gaussian. Specifically, we generate the outcomes using the same models considered in scenario 1 but replace the Gaussian error with $\epsilon_{ki}-0.5\sim \mathrm{beta}(0.5,0.5)$, $k=1,\dots,6$.

\noindent \textbf{Scenario 3 (Non-linear models with Gaussian errors).} This scenario contains the most general case where the time-varying potential compliances are allowed to depend on time-varying covariates (i.e., $X_{11}$) and the outcome model is considered to be non-linear.
The marginals of $D_1$, $D_2$, $D_3$ and $D_4$ are truncated normal with means $0.5X_{01}+0.5X_{02}$, $0.5X_{02}$, $1.5X_{11}-0.5X_{01}$, $1.5X_{11}-0.5X_{02}$ respectively and variance 0.25. For joint distribution of the potential compliances, we  use a Gaussian copula with a 4-dimensional correlation matrix 
\begin{align*}
R^\star=
    \begin{pmatrix}
    1 & 0.2 & 0.2 & 0.2\\
    0.2 & 1 & 0.2 & 0.2\\
    0.2 & 0.2 & 1 & 0.2\\
    0.2 & 0.2 & 0.2 & 1
    \end{pmatrix}
    .
\end{align*}
The potential outcomes are generated as below:
\begin{align*}
    Y_{1i}=& 0.7+0.6\exp{(1+D_{1i})}+0.8X_{01i}-0.2X_{02i}+\epsilon_{1i},\\ 
    Y_{2i}=& 0.2+0.7D_{1i}+0.7D_{2i}+0.9D_{3i}-0.9X_{01i}+0.3X_{02i}+0.7X_{11i}+\epsilon_{2i},\\
    Y_{3i}=& 0.2+ 0.6D_{1i}+0.7D_{2i}+0.8D_{3i}+0.9X_{01i}+0.2X_{2i}+0.6X_{11i}+\epsilon_{3i},\\
    Y_{4i}=& 0.7+0.6D_{1i}+0.6D_{2i}+0.8X_{01i}-0.2X_{02i}+\epsilon_{4i},\\
    Y_{5i}=& 0.3+0.5D_{1i}+0.6D_{2i}+0.7\log(1+D_{4i})-0.5X_{02i}+X_{11i}+\epsilon_{5i},\\
    Y_{6i}=& 0.3+0.8D_{1i}+0.7D_{2i}+0.3D_{4i}-0.5X_{02i}+0.9X_{11i}+\epsilon_{6i},
\end{align*}
with $\epsilon_{ki}\sim \mathrm{N}(0,0.1^2)$, $k=1,\dots,6$.

\subsection{Results}

We compare the performance of our proposed non-parametric Bayes approach with the semiparametric Bayes method of \cite{artman2020adjusting}. Tables \ref{tabs1}-\ref{tab7} summarize the results corresponding to scenarios 1-3, respectively. Note that the results in these tables are multiplied by 10. In scenario 1 (Table \ref{tabs1}), as expected, the semiparametric approach performs well when the outcome models are correctly specified. However, the proposed approach still shows smaller bias and standard errors across all the sample sizes considered. Specifically, the proposed approach results in estimators with up to 10 and 5 times smaller bias and standard errors, respectively.   As the sample size increases the gap in bias values shrinks while the standard errors of the semiparametric estimators remain roughly two times higher than the proposed estimator. 

In scenario 2 (Table \ref{tab6}), while the functional form of outcome model is correctly specified in the semiparametric approach, the deviation from the Gaussian error distribution has resulted in considerable bias and inflated standard errors in the  semiparametric estimators. The proposed approach remains unbiased with low standard errors.  In scenario 3 (Table \ref{tab7}), we fit a linear outcome model in the semiparametric approach which results in a misspecified model. The misspecification induced bias in some of the treatment sequences. For example, in sequences 3 and 4, the  bias terms are 0.418 and 0.889  when $n=250$. The latter biases do not seem to converge to zero as the sample size increases. The proposed method, however, captures the non-linear functional forms, thereby producing unbiased estimators with smaller standard errors relative to the semiparametric approach. 

Along with the treatment sequences, we also study the embedded DTRs for the three scenarios. We use the multiple comparisons with the best (MCB) \citep{hsu1981simultaneous,ertefaie2016identifying} method to identify the EDTRs that are the best, insignificantly different from the best, and significantly different from the best. 
The set of the best EDTRs can be defined as $\mathcal{B}=\{\text{EDTR}_l\vert \text{EDTR}_l\ \text{is not inferior to the best}\}$. To construct such a set, we form simultaneous credible intervals $Y^{(l)}-\max_{l^\prime}Y^{(l^\prime)}$. If the interval contains zero, then it is statistically indistinguishable from the best \citep{artman2020adjusting}. The advantage of using MCB over the pairwise comparisons is that we only have to do $L-1$ comparisons, with $L$ being the number of EDTRs. 

We study four specific compliance levels, namely 100\%, 75\%, 50\% and 25\%. For each level, we compute  the average bias and standard errors for the embedded DTRs for the three scenarios (Table \ref{tabss1}). We also demonstrate the mean outcome for each embedded DTR and the percentage of times they are included in the set of the best EDTRs in Table \ref{tabs2}. For Scenarios 1 and 2, EDTRs 1, 3 and 4 are best for all 200 datasets. For Scenario 3, while EDTRs 1 and 2 are in the set of best EDTRs all the time, EDTRs 3 and 4 are also included a few times.  


\begin{table}[h]
\centering
\small
\caption{Scenario 1.  Bias and standard errors (multiplied by 10) for the conditional mean of outcomes for each  treatment sequence.}
\begin{tabular}{ |c|c|c|c|c|c|c| } 
\hline
& \multicolumn{6}{c|}{Non-parametric Bayes}\\
 \hline
 $n$ & Seq 1 & Seq 2 & Seq 3 & Sec 4 & Seq 5 & Seq 6 \\ 
 \hline
 250 & -0.05(0.83) & 0.04(0.79) & 0.02(0.97) & 0.01(0.86) & 0.01(0.91) & 0.02(0.98)\\
 \hline
 500 & 0.01(0.63) & -0.02(0.47) & 0.02(0.55) & 0.00(0.54) & 0.01(0.48) & 0.01(0.49)\\
 \hline
 1000 & 0.00(0.28) & 0.00(0.32) & 0.00(0.35) & 0.00(0.29) & 0.00(0.25) & 0.00(0.32)\\
 \hline
 & \multicolumn{6}{c|}{Semiparametric Bayes}\\
 \hline
 250 & 0.28(3.48) & 0.01(4.23) & 0.84(4.33) & 0.11(3.07) & -0.11(1.24) & 0.21(1.64)\\
 \hline
 500 & -0.10(0.98) & 0.04(1.05) & 0.06(0.97) & 0.12(0.46) & 0.10(1.19) & -0.20(1.10)\\
 \hline
 1000 & 0.10(0.63) & -0.01(0.65) & 0.03(0.63) & 0.10(0.70) & 0.10(0.54) & 0.02(0.55)\\
 \hline
\end{tabular}
\label{tabs1}
\end{table}

\begin{table}[h]
\centering
\small
\caption{Scenario 2.  Bias and standard errors (multiplied by 10) for the conditional mean of outcomes for each  treatment sequence.}
\begin{tabular}{ |c|c|c|c|c|c|c| } 
\hline
& \multicolumn{6}{c|}{Non-parametric Bayes}\\
 \hline
 $n$ & Seq 1 & Seq 2 & Seq 3 & Sec 4 & Sec 5 & Seq 6 \\ 
  \hline
 250 & 0.05(1.31) & 0.04(0.96) & -0.02(0.90) & 0.01(1.10) & 0.01(0.98) & 0.02(0.82)\\
 \hline
 500 & 0.01(0.64) & 0.02(0.63) & 0.02(0.67) & 0.00(0.59) & -0.01(0.64) & 0.01(0.64)\\
 \hline
 1000 & 0.01(0.37) & 0.01(0.36) & 0.00(0.40) & 0.01(0.40) & 0.01(0.39) & -0.00(0.38)\\
 \hline
 & \multicolumn{6}{c|}{Semiparametric Bayes}\\
 \hline
 250 & -1.34(2.92) & 0.12(5.62) & 1.34(5.80) & 0.53(3.03) & 0.78(5.97) & 0.24(6.23)\\
 \hline
 500 & -0.57(2.54) & 0.29(3.55) & 0.35(3.51) & 0.28(2.60) & 0.07(3.11) & 0.54(3.15)\\
 \hline
 1000 & -0.38(1.42) & 0.44(1.98) & 0.23(2.01) & 0.11(1.51) & 0.01(2.13) & -0.05(2.22)\\
 \hline
\end{tabular}
\label{tab6}
\end{table}

\begin{table}[h]
\centering
\small
\caption{Scenario 3. Bias and standard errors (multiplied by 10) for the conditional mean of outcomes for each  treatment sequence.}
\begin{tabular}{ |c|c|c|c|c|c|c| } 
\hline
& \multicolumn{6}{c|}{Non-parametric Bayes}\\
 \hline
 $n$ & Seq 1 & Seq 2 & Seq 3 & Sec 4 & Sec 5 & Seq 6 \\ 
  \hline
 250 & 0.04(1.61) & 0.20(0.89) & 0.01(1.03) & 0.00(2.01) & 0.02(0.58) & 0.04(0.65)\\
 \hline
 500 & 0.03(0.86) & 0.01(0.49) & 0.01(0.53) & 0.00(0.58) & 0.02(0.41) & 0.00(0.49)\\
 \hline
 1000 & -0.02(0.55) & 0.00(0.29) & 0.00(0.32) & 0.07(0.38) & 0.01(0.25) & -0.01(0.27)\\
 \hline
 & \multicolumn{6}{c|}{Semiparametric Bayes}\\
 \hline
 250 & 0.31(1.26) & 0.31(3.16) & 4.18(3.04) & 8.89(4.88) & 0.08(1.80) & 0.20(1.95)\\
 \hline
 500 & 0.13(0.72) & -0.05(1.88) & 4.03(1.92) & 3.98(1.82) & 0.07(1.29) & 0.03(1.39)\\
 \hline
 1000 & 0.10(0.45) & 0.07(1.15) & 3.99(1.20) & 4.11(1.79) & 0.12(0.82) & 0.19(0.88)\\
 \hline
\end{tabular}
\label{tab7}
\end{table}


\subsection{Sensitivity analysis for Assumption 3}
We examine the sensitivity of our proposed estimator to the violation of Assumption 3. Specifically, in scenario 3, we generate the marginals of the compliances from truncated normal Gaussians as before, but link them together using a $t$-copula with the correlation matrix $R^\star$ instead of a Gaussian copula. All the other aspects of the generative model remains the same. Table \ref{tabs3} shows that the resulting estimators remain unbiased with a slight inflation in the corresponding standard errors compared with the results in Table \ref{tab7}. 
\begin{table}[h]
\centering
\small
\caption{Simulation studies. Bias and standard errors for the four embedded dynamic treatment regimes for the three scenarios discussed in Section 6 of the main paper.}
\begin{tabular}{ |c|c|c|c|c|c| } 
\hline
\multicolumn{6}{c|}{Scenario 1}\\
\hline
 $n$ & Compliance level & EDTR 1 & EDTR 2 & EDTR 3 & EDTR 4 \\ 
 \hline
  \multirow{ 4}{*}{250} & 100\% & -0.07 (0.09) & -0.11 (0.04) & 0.13 (0.06) & 0.10 (0.08)\\
 & 75\% & 0.05 (0.07) & 0.08 (0.05) & -0.08 (0.05) & 0.08 (0.06)\\
 & 50\% & 0.06 (0.05) & 0.07 (0.05) & 0.06 (0.03) & 0.04 (0.04) \\
 & 25\% & 0.06 (0.05) & 0.07 (0.05) & 0.05 (0.06) & 0.03 (0.05)\\
 \hline 
 \multirow{ 4}{*}{500} & 100\% & 0.07 (0.04) & -0.09 (0.04) & 0.11 (0.06) & 0.09 (0.07)\\
 & 75\% & 0.05 (0.06) & 0.06 (0.04) & -0.07 (0.04) & 0.07 (0.05)\\
 & 50\% & 0.05 (0.07) & 0.06 (0.05) & 0.06 (0.04) & 0.05 (0.04)\\
 & 25\% & 0.05 (0.06) & 0.04 (0.05) & 0.04 (0.03) & -0.02 (0.04)\\
 \hline
 \multirow{ 4}{*}{1000} & 100\% & 0.05 (0.04) & -0.07 (0.02) & 0.06 (0.03) & 0.05 (0.05)\\
 & 75\% & 0.04 (0.03) & 0.05 (0.03) & -0.04 (0.03) & 0.04 (0.02)\\
 & 50\% & 0.04 (0.03) & 0.04 (0.05) & 0.03 (0.03) & 0.04 (0.04)\\
 & 25\% & 0.05 (0.03) & 0.03 (0.03) & 0.03 (0.02) & -0.02 (0.03)\\
 \hline
 \multicolumn{6}{c|}{Scenario 2}\\
\hline
 $n$ & Compliance level & EDTR 1 & EDTR 2 & EDTR 3 & EDTR 4\\ 
 \hline
 \multirow{ 4}{*}{250} & 100\% & 0.08 (0.10) & -0.10 (0.06) & 0.11 (0.08) & 0.09 (0.08)\\
 & 75\% & 0.09 (0.08) & 0.09 (0.07) & 0.08 (0.06) & 0.06 (0.07)\\
 & 50\% & 0.06 (0.06) & 0.06 (0.05) & 0.07 (0.05) & 0.05 (0.07)\\
 & 25\% & 0.08 (0.06) & 0.08 (0.07) & 0.07 (0.06) & 0.06 (0.05)\\
 \hline
 \multirow{ 4}{*}{500} & 100\% & -0.05 (0.04) & -0.10 (0.05) & 0.09 (0.07) & 0.08 (0.05)\\
 & 75\% & 0.07 (0.06) & 0.06 (0.07) & 0.07 (0.04) & 0.05 (0.05)\\
 & 50\% & 0.05 (0.04) & 0.08 (0.04) & 0.04 (0.06) & 0.03 (0.04)\\
 & 25\% & 0.07 (0.05) & 0.05 (0.06) & 0.04 (0.05) & 0.04 (0.05)\\
 \hline
 \multirow{ 4}{*}{1000} & 100\% & -0.04 (0.05) & -0.08 (0.04) & 0.07 (0.06) & 0.06 (0.04)\\
 & 75\% & 0.05 (0.06) & 0.04 (0.05) & 0.06 (0.04) & 0.03 (0.05)\\
 & 50\% & 0.03 (0.04) & 0.05 (0.06) & 0.03 (0.02) & 0.02 (0.04)\\
 & 25\% & 0.04 (0.04) & 0.04 (0.05) & 0.03 (0.05) & 0.04 (0.04)\\
 \hline
 \multicolumn{6}{c|}{Scenario 3}\\
\hline
 $n$ & Compliance level & EDTR 1 & EDTR 2 & EDTR 3 & EDTR 4 \\ 
 \hline
 \multirow{4}{*}{250} & 100\% & 0.09(0.10) & 0.12(0.13) & 0.08(0.03) & 0.10(0.06) \\
 & 75\% & 0.08(0.08) & 0.10(0.08) & 0.08(0.07) & 0.08(0.03) \\
 & 50\% & 0.09(0.07) & 0.09(0.09) & 0.09(0.03) & 0.05(0.04) \\
 & 25\% & 0.10 (0.08) & 0.11 (0.05) & 0.10 (0.04) & 0.06 (0.05)\\
 \hline
 \multirow{4}{*}{500} & 100\% & 0.09(0.09) & 0.10(0.11) & 0.07(0.04) & 0.11(0.06) \\
 & 75\% & 0.07(0.04) & 0.08(0.08) & 0.07(0.05) & 0.07(0.03) \\
 & 50\% & 0.08(0.04) & 0.08(0.07) & 0.08(0.04) & 0.04(0.04) \\
 & 25\% & 0.09 (0.05) & 0.07 (0.06) & 0.09 (0.04) & 0.06 (0.03)\\
 \hline
 \multirow{4}{*}{1000} & 100\% & 0.07(0.04) & 0.07(0.07) & 0.05(0.03) & 0.09(0.03) \\
 & 75\% & 0.06(0.04) & 0.06(0.05) & 0.05(0.04) & 0.06(0.03) \\
 & 50\% & 0.05 (0.03) & 0.06(0.05) & 0.08(0.02) & 0.04(0.03) \\
 & 25\% & 0.08 (0.02) & 0.07 (0.02) & 0.05 (0.02) & 0.05 (0.02)\\
 \hline
\end{tabular}
\label{tabss1}
\end{table}

\begin{table}[h]
\centering
\small
\caption{Simulation studies. Mean outcomes for the four embedded dynamic treatment regimes for the three scenarios discussed in Section 6 of the main paper, with the percentage of times they are included in the set of best EDTRs based on 200 datasets (in parentheses).}
\begin{tabular}{ |c|c|c|c|c|c| } 
\hline
\multicolumn{6}{c|}{Scenario 1}\\
\hline
 $n$ & Compliance level & EDTR 1 & EDTR 2 & EDTR 3 & EDTR 4 \\ 
 \hline
 \multirow{4}{*}{250} & 100\% & 1.09 (100) & 0.79 (0) & 1.47 (100) & 1.72 (100)\\
 & 75\% & 0.99 (100) & 0.66 (0) & 1.14 (100) & 1.51 (100)\\
 & 50\% & 0.86 (100) & 0.43 (0) & 1.01 (100) & 1.09 (100)\\
 & 25\% & 0.69 (100) & 0.31 (0) & 0.77 (100) & 0.81 (100)\\
 \hline
 \multirow{4}{*}{500} & 100\% & 1.13 (100) & 0.68 (0) & 1.45 (100) & 1.75 (100)\\
 & 75\% & 1.00 (100) & 0.64 (0) & 1.19 (100) & 1.49 (100)\\
 & 50\% & 0.82 (100) & 0.44 (0) & 1.02 (100) & 1.11 (100)\\
 & 25\% & 0.65 (100) & 0.30 (0) & 0.79 (100) & 0.80 (100)\\
 \hline
 \multirow{4}{*}{1000} & 100\% & 1.12 (100) & 0.71 (0) & 1.46 (100) & 1.80 (100)\\
 & 75\% & 0.98 (100) & 0.61 (0) & 1.21 (100) & 1.47 (100)\\
 & 50\% & 0.80 (100) & 0.44 (0) & 0.99 (100) & 1.11 (100)\\
 & 25\% & 0.64 (100) & 0.32 (0) & 0.80 (100) & 0.83 (100)\\
 \hline
 \multicolumn{6}{c|}{Scenario 2}\\
\hline
 $n$ & Compliance level & EDTR 1 & EDTR 2 & EDTR 3 & EDTR 4\\ 
 \hline
 \multirow{4}{*}{250} & 100\% & 1.13 (100) & 0.76 (0) & 1.50 (100) & 1.74 (100)\\
 & 75\% & 1.05 (100) & 0.63 (0) & 1.20 (100) & 1.47 (100)\\
 & 50\% & 0.89 (100) & 0.57 (0) & 1.03 (100) & 1.14 (100)\\
 & 25\% & 0.67 (100) & 0.33 (0) & 0.81 (100) & 0.82 (100)\\
 \hline
 \multirow{4}{*}{500} & 100\% & 1.14 (100) & 0.74 (0) & 1.49 (100) & 1.76 (100)\\ 
 & 75\% & 1.01 (100) & 0.61 (0) & 1.23 (100) & 1.45 (100)\\
  & 50\% & 0.85 (100) & 0.47 (0) & 1.00 (100) & 1.12 (100) \\
  & 25\% & 0.62 (100) & 0.29 (0) & 1.04 (100) & 0.79 (100)\\
  \hline
  \multirow{4}{*}{1000} & 100\% & 1.11 (100) & 0.80 (0) & 1.46 (100) & 1.78 (100)\\
  & 75\% & 0.99 (100) & 0.64 (0) & 1.17 (100) & 1.46 (100)\\
  & 50\% & 0.82 (100) & 0.49 (0) & 1.05 (100) & 1.15 (100)\\
  & 25\% & 0.65 (100) & 0.34 (0) & 0.75 (100) & 0.82 (100)\\
 \hline
 \multicolumn{6}{c|}{Scenario 3}\\
\hline
 $n$ & Compliance level & EDTR 1 & EDTR 2 & EDTR 3 & EDTR 4 \\ 
 \hline
 \multirow{4}{*}{250} & 100\% & 3.93 (100) & 3.39 (100) & 1.85 (5) & 1.80 (0) \\
 & 75\% & 3.19 (100) & 2.74 (100) & 1.57 (9) & 1.55 (8.5) \\
 & 50\% & 2.47 (100) & 1.99 (100) & 1.19 (9.5) & 1.20 (10.5) \\
 & 25\% & 2.21 (100) & 1.67 (100) & 0.84 (7) & 0.81 (6)\\
 \hline
 \multirow{4}{*}{500} & 100\% & 3.93 (100) & 3.41 (100) & 1.83 (3) & 1.82 (0.5) \\
 & 75\% & 3.21 (100) & 2.74 (100) & 1.55 (7) & 1.55 (8) \\
 & 50\% & 2.47 (100) & 2.01 (100) & 1.17 (9) & 1.19 (9.5) \\
 & 25\% & 2.22 (100) & 1.65 (100) & 0.86 (7) & 0.81 (7)\\
 \hline
 \multirow{4}{*}{1000} & 100\% & 3.99 (100) & 3.42 (100) & 1.86 (3.5) & 1.82 (0.5) \\
 & 75\% & 3.23 (100) & 2.77 (100) & 1.52 (5)  & 1.54 (7) \\
 & 50\% & 2.50 (100) & 2.01 (100) & 1.20 (5) & 1.21 (4.5) \\
 & 25\% & 2.24 (100) & 1.64 (100) & 0.89 (7) & 0.84 (6)\\
 \hline
\end{tabular}
\label{tabs2}
\end{table}

\begin{table}[h]
\centering
\small
\caption{Sensitivity analysis of Assumption 3. Bias and standard errors (multiplied by 10) for the conditional mean of outcomes for each  treatment sequence with the Gaussian copula replaced by a $t$-copula.}
\begin{tabular}{ |c|c|c|c|c|c|c| } 
\hline
 $n$ & Seq 1 & Seq 2 & Seq 3 & Sec 4 & Sec 5 & Seq 6 \\ 
  \hline
 250 & 0.25(1.71) & -0.20(1.64) & -0.11(1.05) & -0.05(2.22) & -0.03(1.33) & -0.10(2.10)\\
 \hline
 500 & 0.05(1.04) & -0.08(1.11) & -0.05(0.88) & -0.03(1.86) & -0.02(0.98) & 0.10(1.25)\\
 \hline
 1000 & -0.01(0.57) & -0.00(0.95) & -0.01(0.74) & -0.02(1.01) & -0.01(0.79) & -0.08(0.86)\\
 \hline
\end{tabular}
\label{tabs3}
\end{table}


\section{Real data application: ENGAGE study}
\label{s:real}

\begin{figure}[t]
    \centering
    \scalebox{0.8}{\includegraphics{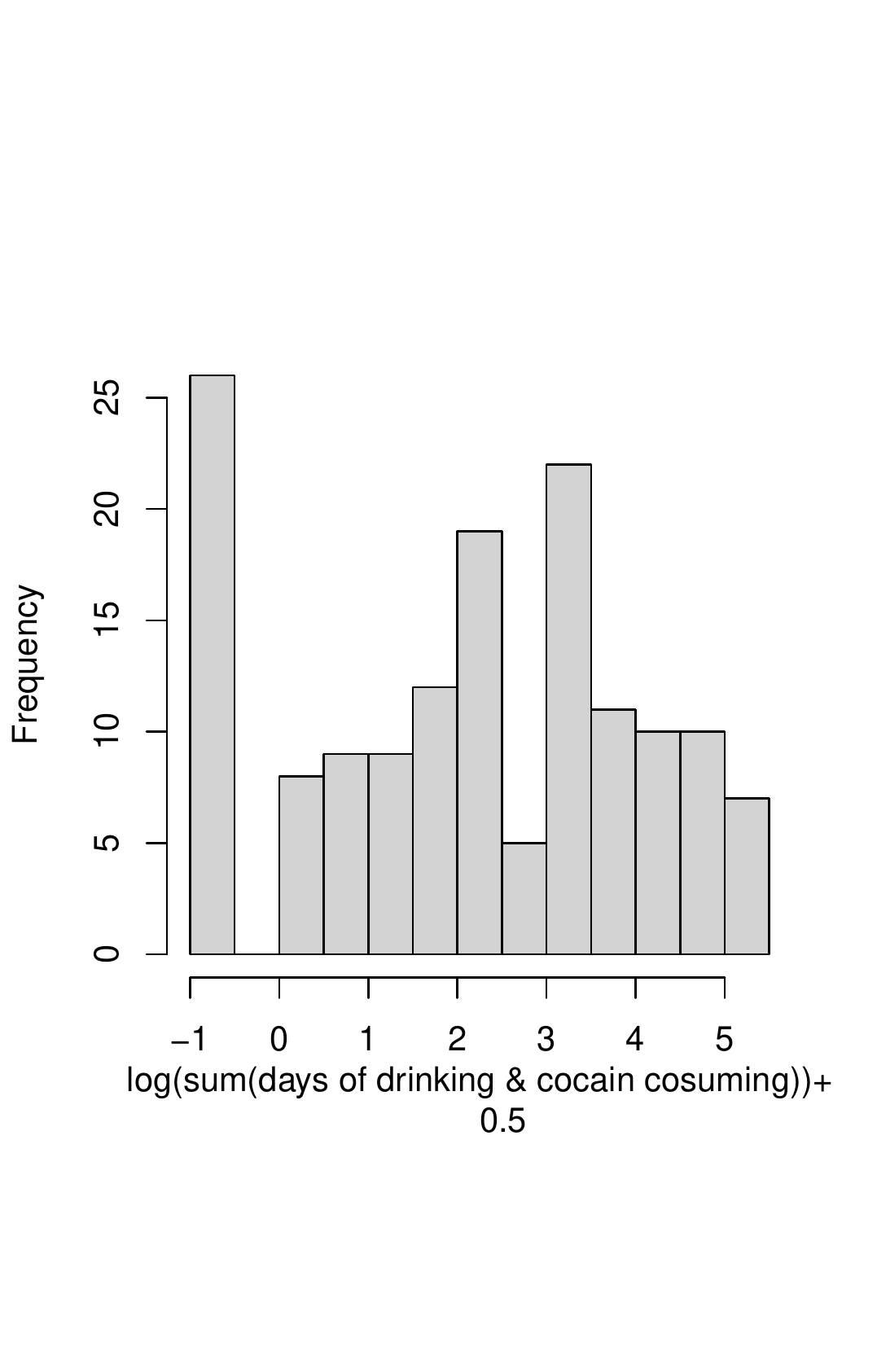}}
    \caption{ENGAGE study. Histogram of log of sum of days of drinking and sum of days of consuming cocaine}
    \label{fig1}
\end{figure}

In this section, we apply our proposed method to the ENGAGE study, and determine which EDTRs are best for certain compliance levels. The data provides longitudinal measurements on patients' ongoing performance given time-varying covariates. We define the outcome as the log of the sum of days from weeks 2 to 24 in which the patient consumed alcohol and the sum of days the patient consumed cocaine plus a small positive constant (0.5), which indicates that smaller values correspond to better outcomes. The sample size is $n=148$, and the covariates include gender (male and female), race (African-American, and non-African-American), and education level. In Figure \ref{fig1}, we plot the histogram of the observed outcomes, and it exhibits a bimodal shape. Hence the parametric assumption of a normal family may be too restrictive which highlights the importance of modeling the potential outcomes using a non-parametric Bayesian approach.
The effect of interventions for substance use varies across gender and race \citep{verissimo2017influence}. We  perform our analyses on the following three subgroups: African-American men, non-African-American men, and African-American women. The subgroup non-African-American women is omitted due to the small sample size. We study four specific compliance values: $100\%$, $75\%$, $50\%$ and $25\%$. For each stratum, we set the education level to the median as well.
\begin{table}[h]
\centering
\small
\caption{ENGAGE study. Point estimates of average potential outcomes along with the standard errors for four embedded dynamic treatment regimes given different compliance levels (i.e., PCEs)}
\begin{tabular}{cccccc}
\multicolumn{5}{c}{Mean EDTR outcomes for Afro-American Men}\\
\hline\\
Compliance level & EDTR 1 & EDTR 2 & EDTR 3 & EDTR 4 & best EDTR
\\  
\hline
100\% & 1.31 (0.22) & 1.28 (0.19) & 2.01 (0.21) & 1.98 (0.21) & 1,2\\
\hline
75\% & 1.39 (0.22) & 1.34 (0.21) & 2.69 (0.21) & 2.21 (0.21) & 1,2\\
\hline
50\% & 1.65 (0.25) & 1.61 (0.22) & 2.89 (0.18) & 2.29 (0.22) & 1,2\\
\hline
25\% & 1.83 (0.30) & 1.99 (0.25) & 3.14 (0.20) & 2.30 (0.24) & 1,2,4\\
\hline
\multicolumn{5}{c}{Mean EDTR outcomes for non Afro-American men}\\
\hline\\
Compliance level & EDTR 1 & EDTR 2 & EDTR 3 & EDTR 4 & best EDTR
\\  
\hline
100\% & 1.07 (0.21) & 1.09 (0.20) & 2.26 (0.21) & 1.88 (0.21) & 1,2\\
\hline
75\% & 1.20 (0.23) & 1.29 (0.19) & 2.35 (0.21) & 1.87 (0.22) & 1,2\\
\hline
50\% & 1.88 (0.27) & 1.84 (0.26) & 2.70 (0.19) & 2.05 (0.24) & 1,2\\
\hline
25\% & 2.15 (0.32) & 2.13 (0.28) & 3.05 (0.23) & 2.17 (0.27) & 1,2,4\\
\hline
\multicolumn{5}{c}{Mean EDTR outcomes for Afro-American women}\\
\hline\\
Compliance level & EDTR 1 & EDTR 2 & EDTR 3 & EDTR 4 & best EDTR
\\  
\hline
100\% & 1.36 (0.21) & 1.21 (0.21) & 2.24 (0.22) & 1.97 (0.23) & 1,2\\
\hline
75\% & 1.38 (0.23) & 1.31 (0.22) & 2.55 (0.18) & 2.06 (0.16) & 1,2\\
\hline
50\% & 1.82 (0.21) & 1.68 (0.19) & 2.64 (0.19) & 2.19 (0.21) & 1,2\\
\hline
25\% & 2.21 (0.29) & 2.08 (0.25) & 3.01 (0.19) & 2.44 (0.24) & 1,2,4\\
\hline
\multicolumn{5}{c}{Mean EDTR outcomes for the overall population}\\
\hline\\
Compliance level & EDTR 1 & EDTR 2 & EDTR 3 & EDTR 4 & best EDTR
\\  
\hline
100\% & 1.21 (0.21) & 1.12 (0.21) & 2.13 (0.22) & 1.91 (0.21) & 1,2\\
\hline
75\% & 1.25 (0.22) & 1.30 (0.21) & 2.41 (0.19) & 2.01 (0.19) & 1,2\\
\hline
50\% & 1.72 (0.24) & 1.77 (0.22) & 2.71 (0.19) & 1.98 (0.21) & 1,2\\
\hline
25\% & 1.98 (0.29) & 2.06 (0.26) & 3.06 (0.20) & 2.21 (0.25) & 1,2,4\\
\hline
\end{tabular}
\label{tabdtr1}
\end{table}
In Table \ref{tabdtr1}, we summarize the mean EDTR outcomes along with the standard errors for the overall population, as well as the different subgroups. We use the multiple comparisons with the best (MCB) \citep{hsu1981simultaneous,ertefaie2016identifying} method to identify the EDTRs that are the best, insignificantly different from the best, and significantly different from the best. 

We again use the MCB method to obtain a credible interval for the difference between each level mean and the best of the remaining level means. This will allow the clinician to select an optimal EDTR while taking into account the cost etc. Table \ref{tabdtr1} shows that EDTRs 1 and 2 are very close and perform better than EDTRs 3 and 4 across all the compliance levels, except for 25\%. The set of best EDTRs includes EDTR 1 and 2 for the upper level compliances, i.e., 100\%, 75\% and 50\%. For 25\% compliance, the set of best EDTRs consists of EDTRs 1, 2 and 4. That is, patients who exhibit a relatively higher level of compliance benefit from starting with a more stringent intervention MI+IOP. For patients with a lower level of compliance, starting with MI-IOP is not distinguishable from Starting with MI-PC and receiving no further care after. Also, among the subgroups considered, non-African American men have the lowest  number of days with alcohol or cocaine use (in log scale) across different EDTRs and compliance levels.  As expected, as the compliance levels decrease, the mean outcomes under  EDTRs increases suggesting that the intervention effect diminishes by lowering the compliance levels.  

\section{Discussion}
\label{s:discussion}
Accounting for compliance is challenging in clinical trials, since they are post-treatment variables. The notion of potential compliance is useful in this context, as we can determine the embedded DTRs as functions of potential compliances. However, for each treatment sequence, some of the potential compliances are latent, and need to be imputed. Using the proposed Bayesian approach, we treat the missing compliances as unknown parameters. Imputation of the missing compliances and estimation of the potential compliances are tied together, and are done sequentially through the MCMC steps. We use Metropolis-Hastings steps embedded inside a Block Gibbs sampler. 

A nonparametric approach for modeling the potential compliances as well as the potential outcomes allows us to handle complex underlying distributions, without imposing any parametric assumptions. By adapting this flexible nonparametric approach, we are able to achieve small bias and standard errors across different types of generative models and error distributions. An interesting extension of our proposed method will be to use a nonparametric copula instead of a Gaussian copula, although it might make the posterior computations a little more complex. Another important extension will be to investigate the theoretical properties of the proposed model. Establishing the large sample properties (for example, posterior consistency) for the conditional expectations $E(Y\vert \bm{D},X_0)$ is challenging due to the randomness of the imputed $\bm{D}$'s. However, since Dirichlet process mixtures have attractive theoretical properties, one can expect to have at least posterior consistency, but the details merit further research.





%
\bibliographystyle{apalike}
\bibliography{mybiblio.bib}

\begin{thebibliography}{}

\bibitem[Artman et~al., 2020]{artman2020adjusting}
Artman, W.~J., Ertefaie, A., Lynch, K.~G., McKay, J.~R., and Johnson, B.~A.
  (2020).
\newblock Adjusting for partial compliance in smarts: a bayesian semiparametric
  approach.
\newblock {\em arXiv preprint arXiv:2005.10307}.

\bibitem[Barnard et~al., 2000]{barnard2000modeling}
Barnard, J., McCulloch, R., and Meng, X.-L. (2000).
\newblock Modeling covariance matrices in terms of standard deviations and
  correlations, with application to shrinkage.
\newblock {\em Statistica Sinica}, pages 1281--1311.

\bibitem[Bartolucci and Grilli, 2011]{bartolucci2011modeling}
Bartolucci, F. and Grilli, L. (2011).
\newblock Modeling partial compliance through copulas in a principal
  stratification framework.
\newblock {\em Journal of the American Statistical Association},
  106(494):469--479.

\bibitem[Bembom and van~der Laan, 2008]{bembom2008analyzing}
Bembom, O. and van~der Laan, M.~J. (2008).
\newblock Analyzing sequentially randomized trials based on causal effect
  models for realistic individualized treatment rules.
\newblock {\em Statistics in {M}edicine}, 27(19):3689--3716.

\bibitem[Chakraborty, 2013]{chakraborty2013statistical}
Chakraborty, B. (2013).
\newblock {\em Statistical methods for dynamic treatment regimes}.
\newblock Springer.

\bibitem[Cheng and Small, 2006]{cheng2006bounds}
Cheng, J. and Small, D.~S. (2006).
\newblock Bounds on causal effects in three-arm trials with non-compliance.
\newblock {\em Journal of the Royal Statistical Society: Series B (Statistical
  Methodology)}, 68(5):815--836.

\bibitem[Cui and Tchetgen~Tchetgen, 2020]{cui2020semiparametric}
Cui, Y. and Tchetgen~Tchetgen, E. (2020).
\newblock A semiparametric instrumental variable approach to optimal treatment
  regimes under endogeneity.
\newblock {\em Journal of the American Statistical Association}, 116:1--12.

\bibitem[Ertefaie et~al., 2016a]{ertefaie2016selection}
Ertefaie, A., Small, D., Flory, J., and Hennessy, S. (2016a).
\newblock Selection bias when using instrumental variable methods to compare
  two treatments but more than two treatments are available.
\newblock {\em The international journal of biostatistics}, 12(1):219--232.

\bibitem[Ertefaie et~al., 2016b]{ertefaie2016sensitivity}
Ertefaie, A., Small, D., Flory, J., and Hennessy, S. (2016b).
\newblock A sensitivity analysis to assess bias due to selecting subjects based
  on treatment received.
\newblock {\em Epidemiology}, 27(2):e5--e7.

\bibitem[Ertefaie et~al., 2016c]{ertefaie2016identifying}
Ertefaie, A., Wu, T., Lynch, K.~G., and Nahum-Shani, I. (2016c).
\newblock Identifying a set that contains the best dynamic treatment regimes.
\newblock {\em Biostatistics}, 17(1):135--148.

\bibitem[Frangakis and Rubin, 2002]{frangakis2002principal}
Frangakis, C.~E. and Rubin, D.~B. (2002).
\newblock Principal stratification in causal inference.
\newblock {\em Biometrics}, 58(1):21--29.

\bibitem[Hsu, 1981]{hsu1981simultaneous}
Hsu, J.~C. (1981).
\newblock Simultaneous confidence intervals for all distances from the
  \enquote{best}.
\newblock {\em The Annals of Statistics}, 9:1026--1034.

\bibitem[Ishwaran and James, 2001]{ishwaran2001gibbs}
Ishwaran, H. and James, L.~F. (2001).
\newblock Gibbs sampling methods for stick-breaking priors.
\newblock {\em Journal of the American Statistical Association},
  96(453):161--173.

\bibitem[Kim et~al., 2019]{kim2019bayesian}
Kim, C., Zigler, C.~M., Daniels, M.~J., Choirat, C., and Roy, J.~A. (2019).
\newblock Bayesian longitudinal causal inference in the analysis of the public
  health impact of pollutant emissions.
\newblock {\em arXiv preprint arXiv:1901.00908}.

\bibitem[Kosorok and Moodie, 2015]{kosorok2015adaptive}
Kosorok, M.~R. and Moodie, E.~E. (2015).
\newblock {\em Adaptive treatment strategies in practice: planning trials and
  analyzing data for personalized medicine}, volume~21.
\newblock SIAM.

\bibitem[Lin et~al., 2008]{lin2008longitudinal}
Lin, J.~Y., Ten~Have, T.~R., and Elliott, M.~R. (2008).
\newblock Longitudinal nested compliance class model in the presence of
  time-varying noncompliance.
\newblock {\em Journal of the American Statistical Association},
  103(482):462--473.

\bibitem[Luedtke and Van Der~Laan, 2016]{luedtke2016statistical}
Luedtke, A.~R. and Van Der~Laan, M.~J. (2016).
\newblock Statistical inference for the mean outcome under a possibly
  non-unique optimal treatment strategy.
\newblock {\em Annals of statistics}, 44(2):713.

\bibitem[McKay et~al., 2015]{mckay2015effect}
McKay, J.~R., Drapkin, M.~L., Van~Horn, D.~H., Lynch, K.~G., Oslin, D.~W.,
  DePhilippis, D., Ivey, M., and Cacciola, J.~S. (2015).
\newblock Effect of patient choice in an adaptive sequential randomization
  trial of treatment for alcohol and cocaine dependence.
\newblock {\em Journal of consulting and clinical psychology}, 83(6):1021.

\bibitem[Murphy et~al., 2001]{murphy2001marginal}
Murphy, S.~A., van~der Laan, M.~J., Robins, J.~M., and Group, C. P. P.~R.
  (2001).
\newblock Marginal mean models for dynamic regimes.
\newblock {\em Journal of the American Statistical Association},
  96(456):1410--1423.

\bibitem[Orellana et~al., 2010]{orellana2010dynamic}
Orellana, L., Rotnitzky, A., and Robins, J.~M. (2010).
\newblock Dynamic regime marginal structural mean models for estimation of
  optimal dynamic treatment regimes, part i: main content.
\newblock {\em The international journal of biostatistics}, 6(2):Article--8.

\bibitem[Qi et~al., 2021]{qi2021proximal}
Qi, Z., Miao, R., and Zhang, X. (2021).
\newblock Proximal learning for individualized treatment regimes under
  unmeasured confounding.
\newblock {\em arXiv preprint arXiv:2105.01187}.

\bibitem[Qiu et~al., 2020]{qiu2020optimal}
Qiu, H., Carone, M., Sadikova, E., Petukhova, M., Kessler, R.~C., and Luedtke,
  A. (2020).
\newblock Optimal individualized decision rules using instrumental variable
  methods.
\newblock {\em Journal of the American Statistical Association}, 116:1--18.

\bibitem[Robins, 1986]{robins1986new}
Robins, J. (1986).
\newblock A new approach to causal inference in mortality studies with a
  sustained exposure period—application to control of the healthy worker
  survivor effect.
\newblock {\em Mathematical modelling}, 7(9-12):1393--1512.

\bibitem[Robins et~al., 2008]{robins2008estimation}
Robins, J., Orellana, L., and Rotnitzky, A. (2008).
\newblock Estimation and extrapolation of optimal treatment and testing
  strategies.
\newblock {\em Statistics in {M}edicine}, 27(23):4678--4721.

\bibitem[Robins, 1997]{robins1997causal}
Robins, J.~M. (1997).
\newblock Causal inference from complex longitudinal data.
\newblock In {\em Latent variable modeling and applications to causality},
  pages 69--117. Springer, New York, NY.

\bibitem[Robins et~al., 2000]{robins2000marginal}
Robins, J.~M., Hernan, M.~A., and Brumback, B. (2000).
\newblock Marginal structural models and causal inference in epidemiology.
\newblock {\em Epidemiology}, 11(5):550--560.

\bibitem[Robins et~al., 1994]{robins1994estimation}
Robins, J.~M., Rotnitzky, A., and Zhao, L.~P. (1994).
\newblock Estimation of regression coefficients when some regressors are not
  always observed.
\newblock {\em Journal of the American statistical Association},
  89(427):846--866.

\bibitem[Rosenblum and Van Der~Laan, 2010]{rosenblum2010targeted}
Rosenblum, M. and Van Der~Laan, M.~J. (2010).
\newblock Targeted maximum likelihood estimation of the parameter of a marginal
  structural model.
\newblock {\em The international journal of biostatistics}, 6(2):Article--19.

\bibitem[Schwartz et~al., 2011]{schwartz2011bayesian}
Schwartz, S.~L., Li, F., and Mealli, F. (2011).
\newblock A bayesian semiparametric approach to intermediate variables in
  causal inference.
\newblock {\em Journal of the American Statistical Association},
  106(496):1331--1344.

\bibitem[Sj{\"o}lander et~al., 2009]{sjolander2009sensitivity}
Sj{\"o}lander, A., Humphreys, K., Vansteelandt, S., Bellocco, R., and Palmgren,
  J. (2009).
\newblock Sensitivity analysis for principal stratum direct effects, with an
  application to a study of physical activity and coronary heart disease.
\newblock {\em Biometrics}, 65(2):514--520.

\bibitem[Swanson et~al., 2015]{swanson2015selecting}
Swanson, S.~A., Robins, J.~M., Miller, M., and Hern{\'a}n, M.~A. (2015).
\newblock Selecting on treatment: a pervasive form of bias in instrumental
  variable analyses.
\newblock {\em American journal of epidemiology}, 181(3):191--197.

\bibitem[Toh and Hern{\'a}n, 2008]{toh2008causal}
Toh, S. and Hern{\'a}n, M.~A. (2008).
\newblock Causal inference from longitudinal studies with baseline
  randomization.
\newblock {\em The international journal of biostatistics}, 4(1):Article--22.

\bibitem[van~der Laan and Petersen, 2007]{van2007causal}
van~der Laan, M.~J. and Petersen, M.~L. (2007).
\newblock Causal effect models for realistic individualized treatment and
  intention to treat rules.
\newblock {\em The international journal of biostatistics}, 3(1):Article--3.

\bibitem[Verissimo and Grella, 2017]{verissimo2017influence}
Verissimo, A. D.~O. and Grella, C.~E. (2017).
\newblock Influence of gender and race/ethnicity on perceived barriers to
  help-seeking for alcohol or drug problems.
\newblock {\em Journal of substance abuse treatment}, 75:54--61.

\end{thebibliography}





\section*{Acknowledgements}
Research reported in this manuscript was supported by (1) National Institute on Alcohol Abuse and Alcoholism under award number R21AA027571, (2) National Insitute on Drug Abuse under award number R01DA048764, and (3) National Institute of Neurological Disorders and Stroke under award number R61NS120240.

\end{document}